\documentclass[twocolumn,showpacs,preprintnumbers,amsmath,amssymb,prl,floatfix]{revtex4}
\usepackage{graphicx}% Include figure files
\usepackage{dcolumn}% Align table columns on decimal point
\usepackage{bm}% bold math
\usepackage{color}
\begin{document}

\title{
Microscopic Model Calculations for the Magnetization Process of Layered Triangular-Lattice Quantum Antiferromagnets
}

\author{Daisuke Yamamoto$^{1}$}
\author{Giacomo Marmorini$^{2}$}
\author{Ippei Danshita$^{3}$}
\affiliation{
{$^1$Waseda Institute for Advanced Study, Waseda University, Tokyo 169-8050, Japan}
\\
{$^2$Condensed Matter Theory Laboratory, RIKEN, Saitama 351-0198, Japan\\
and Research and Education Center for Natural Sciences, Keio University, Kanagawa 223-8521, Japan}
\\
{$^3$Yukawa Institute for Theoretical Physics, Kyoto University, Kyoto 606-8502, Japan\\
and Computational Condensed Matter Physics Laboratory, RIKEN, Saitama 351-0198, Japan}
}
\date{\today}% It is always \today, today,
             % but any date may be explicitly specified
\begin{abstract}
Magnetization processes of spin-1/2 layered triangular-lattice antiferromagnets (TLAFs) under a magnetic field $H$ are studied by means of a numerical cluster mean-field method with a scaling scheme. We find that small antiferromagnetic couplings between the layers give rise to several types of extra quantum phase transitions among different high-field coplanar phases. Especially, a field-induced first-order transition is found to occur at $H\approx 0.7 H_s$, where $H_s$ is the saturation field, as another common quantum effect of ideal TLAFs in addition to the well-established one-third plateau. Our microscopic model calculation with appropriate parameters show excellent agreement with experiments on Ba$_3$CoSb$_2$O$_9$ [T. Susuki $et$ $al$., Phys. Rev. Lett. {\bf 110}, 267201 (2013)]. Given this fact, we suggest that the Co$^{2+}$-based compounds may allow for quantum simulations of intriguing properties of this simple frustrated model, such as quantum criticality and supersolid states.
\end{abstract}
\maketitle

%%%%%%%%%%%%%%%%%%%%%%%%%%%%%%%%%%%%%%%%%%%%%%%%%%%%%%%%%%%%%%%%%%%%%%%%%%%%%%%%%%%%%%%%%%%%%%%%%%%%%%%%%%%%%%%%%%%%%%%%%%%%%%%%%%%%%%%%%%%%%%%%%%%%%%%%%%%%%%
%%                                                                           %%
%% Section I: introduction                                                   %%
%%                                                                           %%
%%%%%%%%%%%%%%%%%%%%%%%%%%%%%%%%%%%%%%%%%%%%%%%%%%%%%%%%%%%%%%%%%%%%%%%%%%%%%%%
%%%%%%%%%%%%%%%%%%%%%%%%%%%%%%%%%%%%%%%%%%%%%%%%%%%%%%%%%%%%%%%%%%%%%%%%%%%%%%%
A well-controlled quantum system is capable of efficiently simulating other quantum systems exhibiting intriguing physical properties that are less understood due to the lack of precise control or direct access in the laboratory. Thirty years after Feynman's proposal~\cite{feynman-82}, this idea of ``quantum simulation'' is becoming a reality with the use of cold atoms~\cite{lewenstein-07,bloch-12}, trapped ions~\cite{blatt-12,schneider-12}, quantum dots~\cite{monousakis-02,byrnes-08}, superconducting circuits~\cite{houck-12}, etc. 
The potential applications widely range from condensed-matter physics (e.g., high-$T_c$ superconductivity) to high-energy physics, quantum chemistry, and cosmology~\cite{georgescu-14}. In the recent years, it has been also proposed to use magnetic insulators as a quantum simulator for, e.g., quantum criticality~\cite{giamarchi-08}, Tomonaga-Luttinger liquid~\cite{ruegg-08,klanjsek-08}, Bose glass~\cite{hong-10,yamada-11}, Higgs mode~\cite{ruegg-08-2}, and Efimov effect~\cite{nishida-13}, taking the advantage of controllability of the relevant parameters by pressure or magnetic fields. Triangular-lattice antiferromagnets (TLAFs) are a promising playground for studying topological phase transitions~\cite{kawamura-84,kawamura-07}, supersolids~\cite{yamamoto-13-1,yamamoto-14}, and the physics of frustration, and for testing numerical methods for two-dimensional (2D) systems with the minus-sign problem~\cite{suzuki-93}. Of particular importance upon the simulation is that the experimental system can be well described by a simple model Hamiltonian. A major problem with TLAFs had been the absence of such ideal compounds that can be compared with the model calculations at a quantitative level.

Recently, the appearance of new TLAF materials comprising magnetic Co$^{2+}$ ions~\cite{shirata-12,zhou-12,susuki-13,koutroulakis-14,lee-14,yokota-14} has changed the situation. Shirata $et$ $al$.~\cite{shirata-12} reported that the magnetization curve of Ba$_3$CoSb$_2$O$_9$ powder seems to show excellent agreement with theoretical calculations on the spin-1/2 Heisenberg model~\cite{honecker-04,yoshikawa-04,farnell-09,sakai-11}, including the one-third quantum magnetization plateau~\cite{nishimori-86,chubukov-91,sebastian-08}. This is owing to the fact that the Co$^{2+}$ ions with an effective spin-1/2 form well-separated layers of regular triangular lattice~\cite{shirata-12}. 
However, even in the almost ideal TLAF, the latest experiments with single crystals~\cite{susuki-13,koutroulakis-14} found an unpredicted magnetization anomaly at a strong magnetic field $H\approx 0.7H_s$ with $H_s$ being the saturation field. In other TLAFs with small spins, such as Cs$_2$CuBr$_4$~\cite{fortune-09}, Ba$_3$NiNb$_2$O$_9$~\cite{hwang-12}, and Ba$_3$CoNb$_2$O$_9$~\cite{lee-14}, the experiments have also found a sign of extra quantum spin states at strong magnetic fields. Although mechanisms for the high-field quantum states are now under active discussion~\cite{yamamoto-14,susuki-13,koutroulakis-14,maryasin-13,starykh-14,sellmann-14}, the conclusive interpretation has not been reached yet due to the lack of direct comparison with the corresponding microscopic model.

\begin{figure}[b]
\includegraphics[scale=0.155]{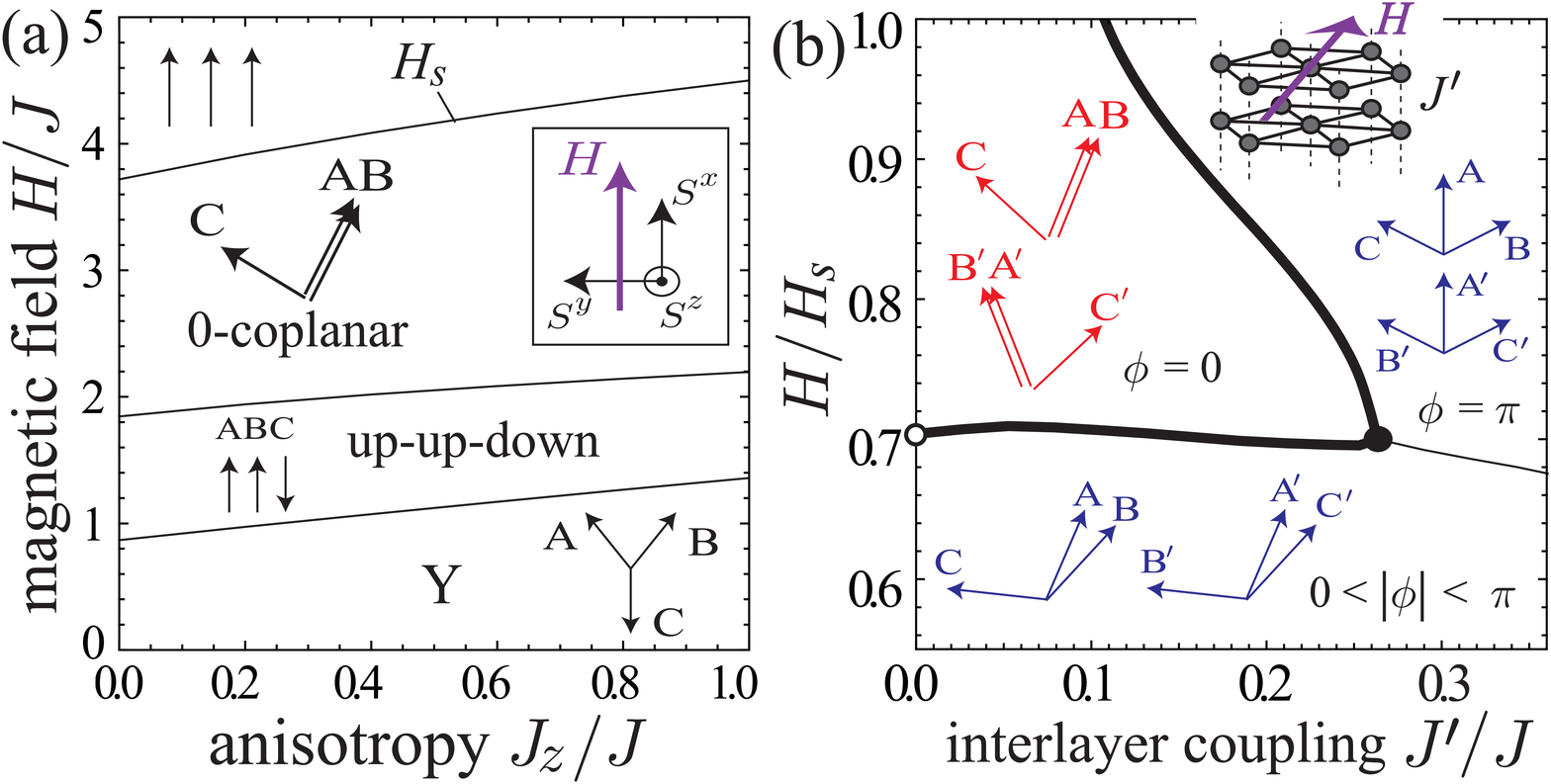}
\caption{\label{fig1}
(color online). (a) Ground-state phase diagram of spin-1/2 easy-plane TLAFs on an independent layer with in-plane magnetic field, obtained by the CMF+S. (b) Emergent high-field states induced by interlayer coupling $J^\prime$ for $J_z/J=0.8$. The thick (thin) curves denote first- (second-) order transitions. }
\end{figure}
In this Letter, we treat a microscopic model of layered TLAFs with the use of a numerical cluster mean-field method with a scaling scheme (CMF+S)~\cite{yamamoto-14,yamamoto-12-1,yamamoto-12-2,yamamoto-13-2,morenocardone-14}.
We quantitatively compare the model calculations with the experiments to provide a microscopic mechanism for the extra high-field quantum states. 
As shown in Fig.~\ref{fig1}(a), the magnetization process of purely 2D TLAFs exhibits three different magnetic phases up to the saturation field $H_s$ with or without easy-plane anisotropy, $0\leq J_z/J\leq 1$. We find that the presence of an antiferromagnetic interlayer coupling $J^\prime$ gives rise to several high-field quantum states shown in Fig.~\ref{fig1}(b). As a result, additional quantum phase transitions between them occur at strong fields even if $J^\prime$ is infinitesimally small. Below, we show the mechanism for the appearance of the high-field states due to quantum fluctuations and the incompatibility between three-sublattice in-plane magnetic orders and the demand of antiparallel alignment along the stacking direction. Moreover, we explain the entire magnetization process of Ba$_3$CoSb$_2$O$_9$ including the high-field magnetization anomaly in terms of the microscopic model in a quantitative way.
The microscopic understanding of the system constitutes a solid foundation for application of Ba$_3$CoSb$_2$O$_9$ as a quantum simulator of TLAF.

%%%%%%%%%%%%%%%%%%%%%%%%%%%%%%%%%%%%%%%%%%%%%%%%%%%%%%%%%%%%%%%%%%%%%%%%%%%%%%%
%%%%%%%%%%%%%%%%%%%%%%%%%%%%%%%%%%%%%%%%%%%%%%%%%%%%%%%%%%%%%%%%%%%%%%%%%%%%%%%
%%                                                                           %%
%% Section II: Model and Method                                              %%
%%                                                                           %%
%%%%%%%%%%%%%%%%%%%%%%%%%%%%%%%%%%%%%%%%%%%%%%%%%%%%%%%%%%%%%%%%%%%%%%%%%%%%%%%
%%%%%%%%%%%%%%%%%%%%%%%%%%%%%%%%%%%%%%%%%%%%%%%%%%%%%%%%%%%%%%%%%%%%%%%%%%%%%%%

%
%%
\begin{figure}[t]
\includegraphics[scale=0.28]{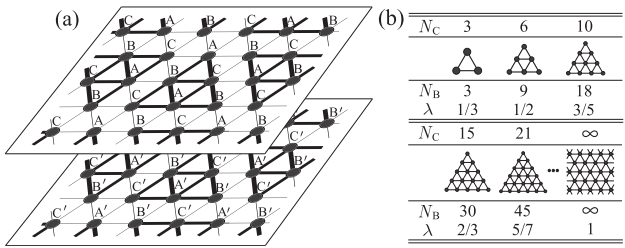}
\caption{\label{fig2}
(a) Six-sublattice structure and the cluster decoupling with $N_C=6$. (b) Series of clusters used in the CMF+S. 
}
\end{figure}
We describe a spin-1/2 layered TLAF with the following model Hamiltonian:
\begin{eqnarray}
\hat{\mathcal{H}}&=&
\sum_{\langle i,j\rangle}\left[J\Big(\hat{S}_i^x\hat{S}_j^x+\hat{S}_i^y\hat{S}_j^y\Big)+J_z\hat{S}_i^z\hat{S}_j^z\right]\nonumber\\
&&+J^\prime\sum_{\langle i,l\rangle^\prime}\hat{\bm{S}}_i\cdot\hat{\bm{S}}_l-H\sum_{i}\hat{S}^x_{i},
\label{hamiltonian}
\end{eqnarray}
where the intralayer ($J,J_z$) and interlayer ($J^\prime$) nearest-neighbor couplings are assumed to be all antiferromagnetic (positive). As well as the isotropic case ($J=J_z$), we consider possible easy-plane anisotropy $0\leq J_z/J< 1$ for the intralayer interactions, which is relevant to many real materials. 
The easy-plane anisotropic system in Eq.~(\ref{hamiltonian}) with the in-plane magnetic field $H$ does not preserve total spins in any direction, which makes the theoretical treatment more challenging. Moreover, the classical limit ($S\rightarrow\infty$) with $J^\prime=0$ exhibits a nontrivial continuous degeneracy of ground states~\cite{kawamura-85,capriotti-98} due to strong frustration. Thus, quantum fluctuations play an essential role in determining the ground-state magnetic ordering as long as the degeneracy lifting term $J^\prime$ is relatively small.

To deal with the quantum effects microscopically, we perform the CMF+S calculations~\cite{yamamoto-14,yamamoto-12-1,yamamoto-12-2,yamamoto-13-2,morenocardone-14}. Under the $3\times 2=6$ sublattice ansatz, the magnetic moment $m^\alpha_\mu=\langle \hat{S}_\mu^\alpha \rangle$ ($\mu=\{A,B,C,A^\prime,B^\prime,C^\prime\}$) is self-consistently determined by diagonalizing the Hamiltonian on a cluster of $N_C$ spins at zero temperature (see Fig.~\ref{fig2})~\cite{EPAPS}. This approach reproduces the classical ground state for $N_C=1$, and allows for a systematic inclusion of non-local fluctuations as $N_C$ increases. We first discuss the microscopic mechanism for the magnetization process of the model (\ref{hamiltonian}) with the minimal $N_C=3$ cluster, and eventually make an extrapolation to the limit of $N_C \rightarrow \infty$, where long-range fluctuations in each layer are fully included. The scaling parameter $\lambda \equiv N_B/3N_C$ ($N_B$ is the number of bonds treated exactly) varies from 0 for $N_C=1$ to 1 for $N_C=\infty$. The data for each $N_C>3$ and technical details are presented in the Supplementary Material~\cite{EPAPS}.

%%%%%%%%%%%%%%%%%%%%%%%%%%%%%%%%%%%%%%%%%%%%%%%%%%%%%%%%%%%%%%%%%%%%%%%%%%%%%%%
%%%%%%%%%%%%%%%%%%%%%%%%%%%%%%%%%%%%%%%%%%%%%%%%%%%%%%%%%%%%%%%%%%%%%%%%%%%%%%%
%%                                                                           %%
%% Section III: CMF3                                                         %%
%%                                                                           %%
%%%%%%%%%%%%%%%%%%%%%%%%%%%%%%%%%%%%%%%%%%%%%%%%%%%%%%%%%%%%%%%%%%%%%%%%%%%%%%%
%%%%%%%%%%%%%%%%%%%%%%%%%%%%%%%%%%%%%%%%%%%%%%%%%%%%%%%%%%%%%%%%%%%%%%%%%%%%%%%
%
%%
\begin{figure}[t]
\includegraphics[scale=0.285]{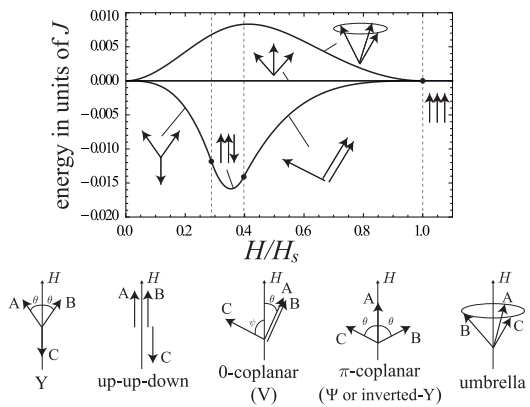}
\caption{\label{fig3}
Energy per site of candidate magnetic orders measured from that of the $\pi$-coplanar state for $J=J_z$ and $J^\prime =0$ within the $N_C=3$ approximation. The lower illustrations depict the spin configurations on an elementary triangle ABC. 
}
\end{figure}
Figure~\ref{fig3} shows the $N_C=3$ result for the energies of candidate magnetization processes of an isotropic 2D system ($J=J_z$ and $J^\prime =0$). We introduce the parameter $\phi\equiv 6{\rm Arg} \left[m^y_A+m^y_Be^{i 2\pi/3}+m^y_Ce^{i 4\pi/3}\right]$ ($-\pi<\phi\leq \pi$), which takes value 0 for the 0-coplanar state (also called the V state) and $\pi$ for the Y state and the $\pi$-coplanar state (also called the $\Psi$ state~\cite{starykh-14} or the inverted-Y state). The quantum effects~\cite{chubukov-91} select the sequence of the Y, up-up-down, and 0-coplanar states for the ground state below $H_{s}=9J/2$~\cite{honecker-04,yoshikawa-04,farnell-09,sakai-11}. The purely 2D model appears to be a good approximation for TLAFs with well-separated magnetic layers~\cite{shirata-12,zhou-12,susuki-13,koutroulakis-14}. However, we will show that even quite a small interlayer coupling can give rise to another quantum phase transition that is absent in the 2D model.

\begin{figure}[t]
\includegraphics[scale=0.27]{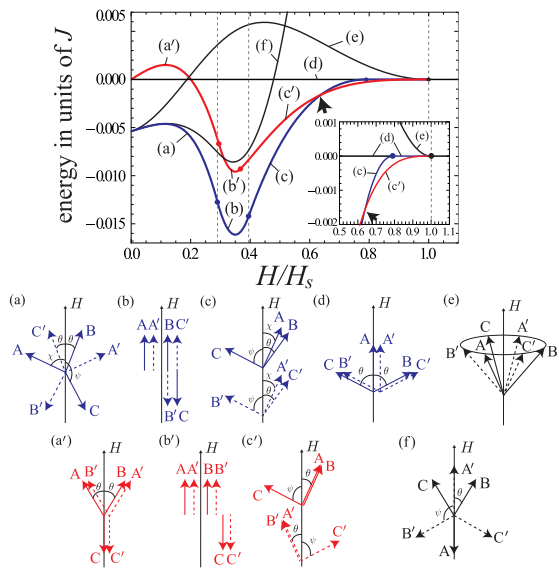}
\caption{\label{fig4} (color online). Energy per site of candidate magnetic orders measured from that of the configuration (d) for $J=J_z$ and $J^\prime =0.025J$ within the $N_C=3$ approximation. The arrow locates a first-order transition that is absent in the 2D result. The inset shows an enlarged view of the high-field region. The lower illustrations depict the spin configurations on an elementary triangular prism ABC-A$^\prime$B$^\prime$C$^\prime$. }
\end{figure}

Figure~\ref{fig4} shows the ground-state selection when an interlayer coupling $J^\prime=0.025J$ exists. The saturation field is now given by $H_{s}=9J/2+2J^\prime$. We see that the sequence of the Y, up-up-down, and 0-coplanar states is separated into two branches, (a)-(b)-(c)-(d) and (a$^\prime$)-(b$^\prime$)-(c$^\prime$). 
At $0.3\lesssim H/H_{s}\lesssim 0.4$, the stacked up-up-down structure of (b) has the lowest energy. Since $J^\prime>0$, the spins tend to align antiparallel with their neighbors in the stacking direction. However, this demand cannot be completely fulfilled for the three-sublattice up-up-down structure in contrast to the standard N\'eel ordering on layered bipartite lattices. Whereas the first two pairs of spins connected by $J^\prime$ can align antiparallel in a unit prism ABC-A$^\prime$B$^\prime$C$^\prime$, the third one has to be oriented in the same direction. When one increases $H$, the configuration (b) is transformed into a stacked coplanar state (c). The geometric incompatibility %mentioned above 
also inhibits simple stacking of stable 0-coplanar order, and the spins have to form an intermediate $0<|\phi|<\pi$ coplanar order identified by three angles $\theta$, $\chi$, and $\psi$. As $H$ increases, the difference between $\theta$ and $\chi$ becomes larger, and the state (c) eventually merges with the alternately stacked $\pi$-coplanar state (d) at a certain magnetic field ($\approx 0.8H_{s}$).

The branch of (a$^\prime$)-(b$^\prime$)-(c$^\prime$) has higher energy at low magnetic fields since the uniformly stacked up-up-down structure of (b$^\prime$) is disfavored by antiferromagnetic $J^\prime$. 
However, the configuration (b$^\prime$) is connected to a simple alternately stacked 0-coplanar state (c$^\prime$), which can reduce the interlayer bond energy of field-transverse spin components. Therefore, as can be seen near the saturation field of Fig.~\ref{fig4}, the energy of the state (c$^\prime$) becomes lower than that of (d). This is consistent with the expectation from the purely 2D result in Fig.~\ref{fig3}, where the 0-coplanar order is more favored by quantum fluctuations than the $\pi$-coplanar order. Consequently, the two energy curves must cross, i.e., an additional first-order transition occurs at some point between the end of the stacked up-up-down state and the saturation field. Indeed, one can see in Fig.~\ref{fig4} that the ground-state magnetization process for very weak $J^\prime>0$ is given as (a)-(b)-(c)-(c$^\prime$).

%%%%%%%%%%%%%%%%%%%%%%%%%%%%%%%%%%%%%%%%%%%%%%%%%%%%%%%%%%%%%%%%%%%%%%%%%%%%%%%
%%%%%%%%%%%%%%%%%%%%%%%%%%%%%%%%%%%%%%%%%%%%%%%%%%%%%%%%%%%%%%%%%%%%%%%%%%%%%%%
%%                                                                           %%
%% Section IV: CMF+S                                                         %%
%%                                                                           %%
%%%%%%%%%%%%%%%%%%%%%%%%%%%%%%%%%%%%%%%%%%%%%%%%%%%%%%%%%%%%%%%%%%%%%%%%%%%%%%%
%%%%%%%%%%%%%%%%%%%%%%%%%%%%%%%%%%%%%%%%%%%%%%%%%%%%%%%%%%%%%%%%%%%%%%%%%%%%%%%

To make a quantitative comparison with experiments, we take into account longer-range spin flactuations and easy-plane anisotropy. In Fig.~\ref{fig1}, we already showed the numerical CMF+S ($N_C\rightarrow \infty$) results. Figure~\ref{fig1}(a) shows that any extra phase transition is induced only by easy-plane anisotropy in 2D. For $J_z/J\neq 1$, the ${\rm U}(1)$ symmetry with respect to the $x$-axis is absent, and the transition from the saturated to 0-coplanar state at $H=H_s$ is accompanied by the spontaneous breaking of $\mathbb{Z}_6$ symmetry under permutation of sublattices and $\pi$ spin rotation about the magnetic field direction~\cite{chubukov-91}. Moreover, since the total spin is no longer a good quantum number, the value of $H_s$ is reduced from the classical one $9J/2+2J^\prime$ and the magnetization plateaus at $M\approx 1/6$ and $M\approx 1/2$ have a slight but finite positive slope.

\begin{figure}[t]
\includegraphics[scale=0.15]{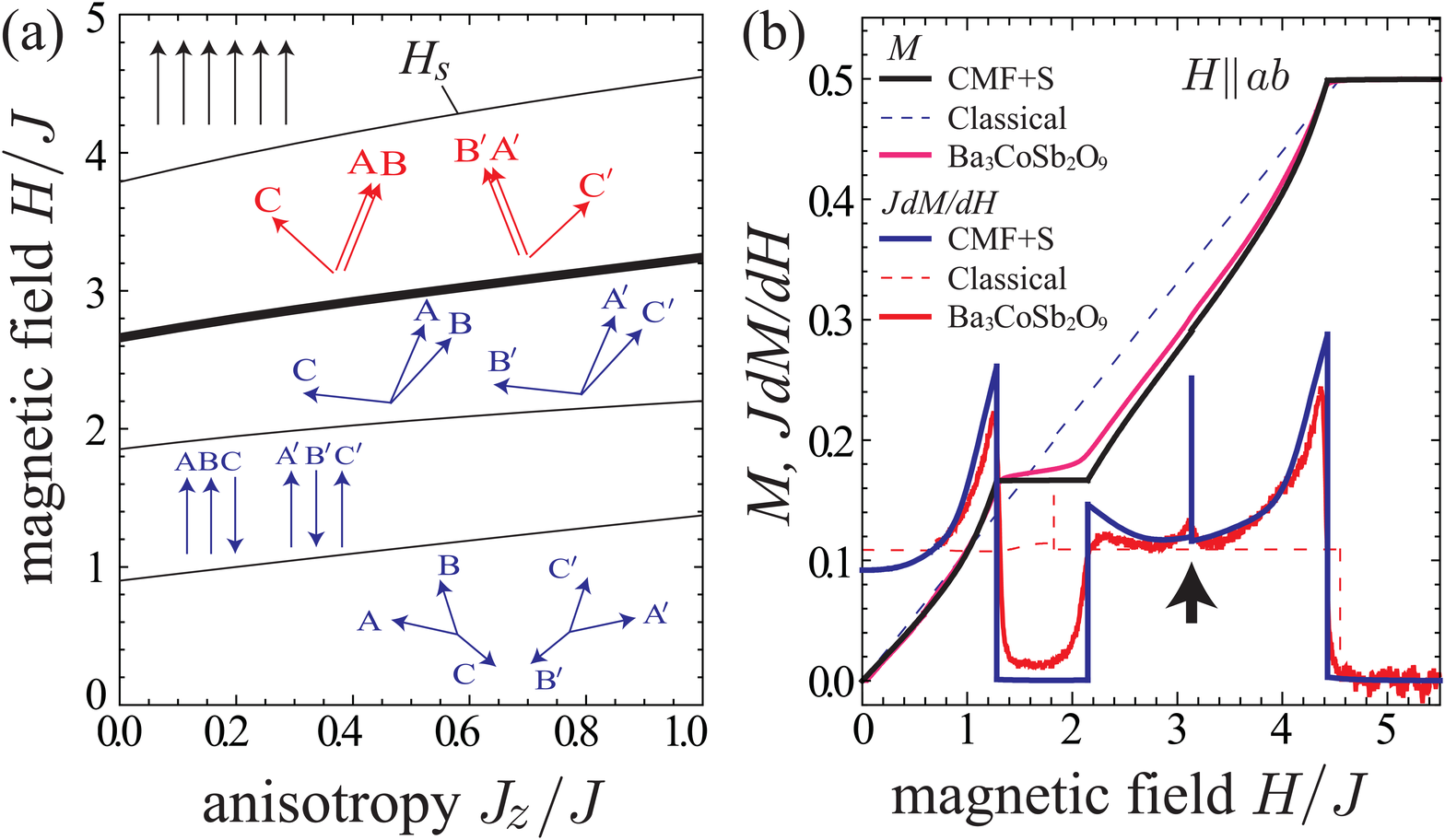}
\caption{\label{fig5} (color online). (a) Ground-state phase diagram for $J^\prime=0.025J$ and $N_C\rightarrow \infty$. The thick (thin) curves denote first- (second-) order transitions. (b) The magnetization curve $M$ and its field derivative $J dM/dH$ for $J^\prime=0.025J$ and $J_z/J=0.8$. The CMF+S at zero temperature is compared to the classical-spin analysis and the experimental data of Ba$_3$CoSb$_2$O$_9$ for $H\parallel ab$ at $T=1.3K\approx 0.07J$~\cite{susuki-13,footnote}. The values of the saturation field and the saturation magnetization are adjusted. 
}
\end{figure}

Figure~\ref{fig1}(b) shows the high-field quantum states induced by interlayer coupling $J^\prime>0$. In the quasi-2D regime ($0<J^\prime\lesssim 0.1J$), the first-order transition between the coplanar states with $0<|\phi|<\pi $ and $\phi=0$ occurs (due to the mechanism we already explained). The first-order transition takes place at $H\approx 0.7H_s$ almost independently of the values of $0<J^\prime/J\lesssim 0.1$ and $J_z/J$ [see also Fig.~\ref{fig5}(a)], which thus can be regarded as a common quantum effect in ideal TLAF compounds, given the unavoidable three dimensionality. For $J^\prime\gtrsim 0.25J$, the second-order transition between the $0<|\phi|<\pi$ and $\phi=\pi$ coplanar states takes place, i.e., the branch of (a)-(b)-(c)-(d) has the lowest energy up to the saturation field. In the intermediate region, $0.1J\lesssim J^\prime\lesssim 0.25J$, the magnetization process follows (a)-(b)-(c)-(c$^\prime$)-(d) with two first-order transitions. Note that the present CMF+S study based on 2D clusters is more reliable for smaller $J'/J$.

Finally, we present an application to the spin-1/2 easy-plane TLAF Ba$_3$CoSb$_2$O$_9$~\cite{shirata-12,zhou-12,susuki-13,koutroulakis-14}. Although the magnetization process exhibits a considerable dependence on the magnetic field direction, the fitting of the resonance conditions with a semiclassical prediction suggests that the exchange anisotropy is very small~\cite{susuki-13}. Nonetheless, the numerical quantum phase diagram shows that the absence of one-third plateau for $H\parallel c$~\cite{susuki-13,koutroulakis-14} cannot be explained with such a small anisotropy~\cite{yamamoto-14,sellmann-14}. This indicates the importance of taking quantum effects into consideration when determining the model parameters. Here, we use the value of $J^z/J= 0.8$ ($\approx 1/1.3$) according to Ref.~\cite{yamamoto-14}. This compound has a well-separated layered structure, and the magnetization curve shows a single unexpected anomaly at a strong in-plane field $H\parallel ab$. Thus, the interlayer coupling $J^\prime$ should be very weak, at most $\sim 0.1J$ from Fig.~\ref{fig1}(b).

In Fig.~\ref{fig5}(b), we show the comparison of the CMF+S magnetization curve $M(H)\equiv\sum_\mu m_\mu^x/6$ at $J^z/J=0.8$ and $J^\prime/J= 0.025$ with the experimental data of Ba$_3$CoSb$_2$O$_9$~\cite{susuki-13}. The model calculation and experiment show excellent agreement including the nonlinear bending of the curve, the width of the plateau, and the anomaly at $H\approx 0.7H_s$, although the plateau and anomaly are slightly smeared in the experimental data at low but finite temperatures. Especially, we find a good correspondence between the observed anomaly and the point at which $J dM/dH$ exhibits a divergence associated with a small jump in $M$ at the first-order transition between the states (c) and (c$^\prime$). This strongly suggests that the compound Ba$_3$CoSb$_2$O$_9$ can be well described by the simple microscopic model of Eq.~(\ref{hamiltonian}), allowing for the use of Ba$_3$CoSb$_2$O$_9$ as a quantum simulator of TLAF.

Regarding the origin of the high-field magnetization anomaly in Ba$_3$CoSb$_2$O$_9$, several conjectures have been proposed very recently. (i) The authors of the first experimental study~\cite{susuki-13} naively expected that the transition to the $\pi$-coplanar state occurred. According to Fig.~\ref{fig1}(b), this may take place only for $J^\prime\gtrsim 0.25J$. (ii) Maryasin and Zhitomirski~\cite{maryasin-13} also proposed that non magnetic impurities could stabilize a fan-like ($\pi$-coplanar) spin order. (iii) The spins aligning along the field-transverse direction on one sublattice might cause a peak in the susceptibility~\cite{sellmann-14}. (iv) An effective classical model with phenomenological biquadratic coupling predicted the occurrence of a transition between two different ``V-like'' states~\cite{koutroulakis-14}, which are equivalent to the (c) and (c$^\prime$) states. 
The present microscopic study supports the phenomenological conjecture (iv) in the quasi-2D regime $J^\prime\lesssim 0.1J$.

We suggest to use the Ba$_3$CoSb$_2$O$_9$ quantum simulator of TLAF for investigating quantum criticality that has not been analyzable with unbiased numerical methods due to the geometrical frustration. Of particular interest is the transition from saturated to 0-coplanar states at $H=H_s$, with which the $\mathbb{Z}_6$ symmetry breaking is associated. The dynamical exponent for this transition is $z=1$ because the excitation spectrum at the transition point exhibits a gapless and linear dispersion~\cite{EPAPS}. Hence, it is expected that this transition belongs to the same universality class as the one of the ordering transition of the ($D$+1)-dimensional six-state clock model, which is known to be the ($D$+1)-dimensional $XY$ universality class~\cite{blankschtein-84, oshikawa-00, bonnes-11}. Since critical behaviors of the transition temperature $T_c$ to disordered phases are often measured in experiments to identify the universality class~\cite{giamarchi-08}, we here describe that of the present case as $T_c\propto |H-H_s|^{\varphi}$. For the ($D$+1)-dimensional $XY$ universality class, $\varphi=z\nu$ below the upper critical dimension $D_c=3$~\cite{fisher-89}, where $\nu$ is the critical exponent for the correlation length. When $H$ is so close to $H_s$ that $T_c \lesssim 2J'$, the interlayer coupling is relevant and the transition nature is three-dimensional ($D=3$). In this case, $\varphi = 1/2$~\cite{sansone-07}. In contrast, when  $T_c \gtrsim 2J'$, thermal fluctuations decouple the layers such that the transition is two-dimensional and $\varphi = 0.6717(1)$~\cite{campostrini-06}.

%%%%%%%%%%%%%%%%%%%%%%%%%%%%%%%%%%%%%%%%%%%%%%%%%%%%%%%%%%%%%%%%%%%%%%%%%%%%%%%
%%%%%%%%%%%%%%%%%%%%%%%%%%%%%%%%%%%%%%%%%%%%%%%%%%%%%%%%%%%%%%%%%%%%%%%%%%%%%%%
%%                                                                           %%
%% Section VI: summary                                                       %%
%%                                                                           %%
%%%%%%%%%%%%%%%%%%%%%%%%%%%%%%%%%%%%%%%%%%%%%%%%%%%%%%%%%%%%%%%%%%%%%%%%%%%%%%%
%%%%%%%%%%%%%%%%%%%%%%%%%%%%%%%%%%%%%%%%%%%%%%%%%%%%%%%%%%%%%%%%%%%%%%%%%%%%%%%
In conclusion, we have studied the magnetization process of spin-1/2 layered TLAFs, motivated by the recent observations of high-field quantum states~\cite{susuki-13,koutroulakis-14,fortune-09,hwang-12,lee-14}. It was shown that even a small antiferromagnetic coupling between the layers can change the nature of the ground state, giving rise to additional quantum phase transitions at a strong field above the one-third magnetization plateau. 
Our microscopic model with the CMF+S approach provides a quantitative agreement with the magnetization process of quasi-2D TLAFs~\cite{susuki-13,koutroulakis-14}, and properly explains the observed magnetization anomaly as a first-order transition between different high-field quantum states. 
Thanks to its quantitative correspondence to the microscopic model, the system of Ba$_3$CoSb$_2$O$_9$ may be used for quantum simulations of important properties of TLAF, such as quantum criticality and supersolid phases~\cite{yamamoto-13-1, yamamoto-14,wang-09,jiang-09,heidarian-10}.

%%%%%%%%%%%%%%%%%%%%%%%%%%%%%%%%%%%%%%%%%%%%%%%%%%%%%%%%%%%%%%%%%%%%%%%%%%%%%%%
%%%%%%%%%%%%%%%%%%%%%%%%%%%%%%%%%%%%%%%%%%%%%%%%%%%%%%%%%%%%%%%%%%%%%%%%%%%%%%%
%%                                                                           %%
%% Section VII: acknowledgements                                             %%
%%                                                                           %%
%%%%%%%%%%%%%%%%%%%%%%%%%%%%%%%%%%%%%%%%%%%%%%%%%%%%%%%%%%%%%%%%%%%%%%%%%%%%%%%
%%%%%%%%%%%%%%%%%%%%%%%%%%%%%%%%%%%%%%%%%%%%%%%%%%%%%%%%%%%%%%%%%%%%%%%%%%%%%%%
We acknowledge the authors of Ref.~\cite{susuki-13} for sending their experimental data in numerical form. We also thank Hidekazu Tanaka and Yoshitomo Kamiya for useful discussions. This work was supported by KAKENHI Grants from JSPS No. 25800228 (I.D.), No. 25220711 (I.D.), and No. 26800200 (D.Y.).

\onecolumngrid

\newpage 

\subsection{\large Supplementary Material for ``Microscopic Model Calculations for the Magnetization Process of Layered Triangular-Lattice Quantum Antiferromagnets''}
\renewcommand{\thesection}{\Alph{section}}
\renewcommand{\thefigure}{S\arabic{figure}}
\renewcommand{\thetable}{S\Roman{table}}
\setcounter{figure}{0}
\newcommand*{\citenamefont}[1]{#1}
\newcommand*{\bibnamefont}[1]{#1}
\newcommand*{\bibfnamefont}[1]{#1}

\def\bs{{\bf S}}
\def\bk{{\bf k}}
\def\bp{{\bf p}}
\def\bq{{\bf q}}
\def\bQ{{\bf Q}}
\def\b0{{\bf 0}}
\def\br{{\bf r}}
\def\vpa{V^{\parallel}}
\def\vpe{V^{\perp}}
\def\dag{\dagger}
\def\cM{{\cal M}}
\def\bra{\langle}
\def\ket{\rangle}
\def\bbra{\langle\!\langle}
\def\kket{\rangle\!\rangle}
\def\vev#1{\langle{#1}\rangle}
\def\emin{\epsilon_{\rm min}}
\def\non{\nonumber\\}
\renewcommand{\theequation}{S\arabic{equation}}
\renewcommand{\thesection}{\Alph{section}}
\renewcommand{\thefigure}{S\arabic{figure}}
\renewcommand{\thetable}{S\Roman{table}}
\subsection{A. Numerical data for $N_C>3$ and examples of the extrapolation in CMF+S} 
In the CMF+S~\cite{S1}, the Hamiltonian $\hat{\mathcal{H}}$ on $N$ sites is approximated by the sum of $N/N_C$ cluster Hamiltonians with effective mean fields on a cluster of $N_C$ sites through the mean-field decoupling of the inter-cluster interactions: $\hat{S}_{i}^{\alpha} \hat{S}_{j}^\alpha \rightarrow \langle \hat{S}_{i}^{\alpha} \rangle \hat{S}_{j}^\alpha + \langle \hat{S}_{j}^{\alpha} \rangle \hat{S}_{i}^\alpha$ ($\alpha=\{x,y,z\}$). Under the $3\times 2=6$ sublattice ansatz shown in Fig. 2(a) of the main text, we have to consider two independent clusters $C_1$ and $C_2$ for $N_C=3$, 6, 15, and 21 [see Fig.~S1]. For $N_C=10$, there are six independent clusters, $C_1$-$C_6$. The sublattice magnetic moment $m^\alpha_\mu$ ($\mu=\{A,B,C,A^\prime,B^\prime,C^\prime\}$) is given by
\begin{eqnarray}
m^\alpha_\mu=  \frac{1}{N_\mu}\sum_{n}\sum_{i_\mu\in C_n}{\rm Tr}\left( \hat{S}_{i_\mu}^{\alpha}e^{-\beta \hat{\mathcal{H}}_{C_n}}\right)\Big{/} {\rm Tr}( e^{-\beta \hat{\mathcal{H}}_{C_n}}), \label{selfconsistent}
\end{eqnarray}
where $n=1,2,\cdots,M_C$ with $M_C$ being the number of the independent clusters, $N_\mu$ is the number of total sites belonging to the sublattice $\mu$ in the $M_C$ clusters, and $\beta=1/T$ (we take $T\rightarrow 0$ in the present study). The cluster Hamiltonian $\hat{\mathcal{H}}_{C_n}$ includes the expectation values $\langle \hat{S}_{i}^{\alpha} \rangle$ as mean fields to be determined self-consistently. Substituting $m^\alpha_\mu$ into $\langle \hat{S}_{i_\mu}^{\alpha} \rangle$ in $\hat{\mathcal{H}}_{C_n}$, Eq.~(\ref{selfconsistent}) becomes a set of self-consistent equations for $m^\alpha_\mu$. The energy difference between different spin configurations can be estimated by integrating the magnetization curve $M(H)\equiv\sum_\mu m_\mu^x/6$ with respect to $H$ from $H_s$.

Figures S2(i) is the data of the phase diagrams for different $N_C$. Here, we refer to the phase boundaries between the states (a) and (b), between (b) and (c), and between (c) and (c$^\prime$) as $H_{c1}$, $H_{c2}$, and $H_{c3}$, respectively. 
We obtained the CMF+S phase diagram in Fig.~5(a) of the main text by a linear extrapolation $N_C\rightarrow \infty$ ($\lambda\rightarrow 1$) of the phase boundaries calculated with the three largest clusters. See examples of the extrapolation in Fig.~S2(ii).

Figures S3(i) is the data of the magnetization curve for different $N_C$. 
To make the extrapolation $N_C\rightarrow \infty$, we rescale each magnetization curve with respect to $H$ so that the phase boundaries for different $N_C$ locates at the same points, i.e., we extrepolate the data of the magnetization curves at each value of 
\begin{eqnarray*}
h_1&\equiv&\frac{H}{H_{c1}(N_C)}~~{\rm for~the~state~(a)},~~
h_2\equiv\frac{H-H_{c1}(N_C)}{H_{c2}(N_C)-H_{c1}(N_C)}~~{\rm for~(b)},~~
h_3\equiv\frac{H-H_{c2}(N_C)}{H_{c3}(N_C)-H_{c2}(N_C)}~~{\rm for~(c)},\\
h_4&\equiv&\frac{H-H_{c3}(N_C)}{{H}_{s}(N_C)-H_{c3}(N_C)}~~{\rm for~(c^\prime)},~~{\rm and}~~h_5\equiv \frac{H-H_s(N_C)}{H_s(N_C)}~~{\rm for}~H>H_s(N_C). 
\end{eqnarray*}
See examples of the extrapolation in Fig.~S3(ii) to obtain the magnetization curve in Fig.~5(b) of the main text.

\subsection{B. Excitation spectra at $H=H_s$}

The linear spin-wave spectrum is given by
$
\omega({\bm k})= \sqrt{(H-3J(1-\gamma({\bm k})))(H-3(J-J_z\gamma({\bm k})))}
$
for $H\geq H_s$ and $J^\prime=0$. Here, $\gamma({\bm k})\equiv \sum_i \cos {\bm k}\cdot {\bm e}_i/3$ with ${\bm e}_1=(1,0)$, ${\bm e}_1=(1/2,\sqrt{3}/2)$, and ${\bm e}_3=(1/2,-\sqrt{3}/2)$.
Figure S4 shows the spectra at $H=H_s$ for the Heisenberg ($J_z/J=1$) and easy-plane ($J_z/J=0.8$) cases.

\begin{figure}[t]
\includegraphics[scale=0.45]{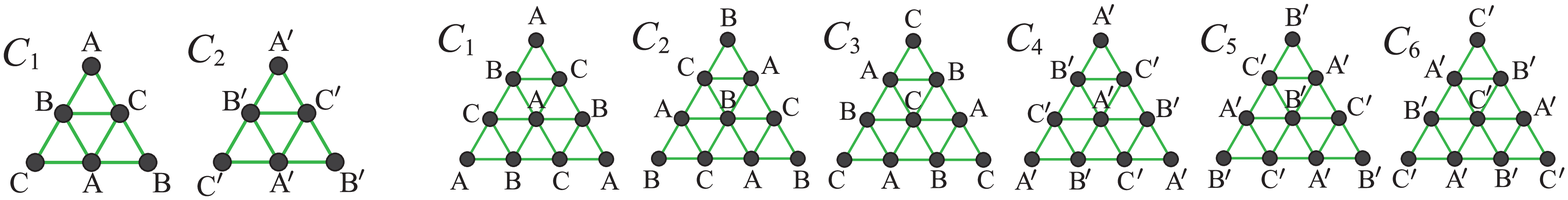}
\caption{\label{figS1}
The clusters that have to be considered in calculations with $N_C=6$ ($M_C=2$) and $N_C=10$ ($M_C=6$), respectively. 
}
\end{figure}
\begin{figure}[t]
\includegraphics[scale=0.3]{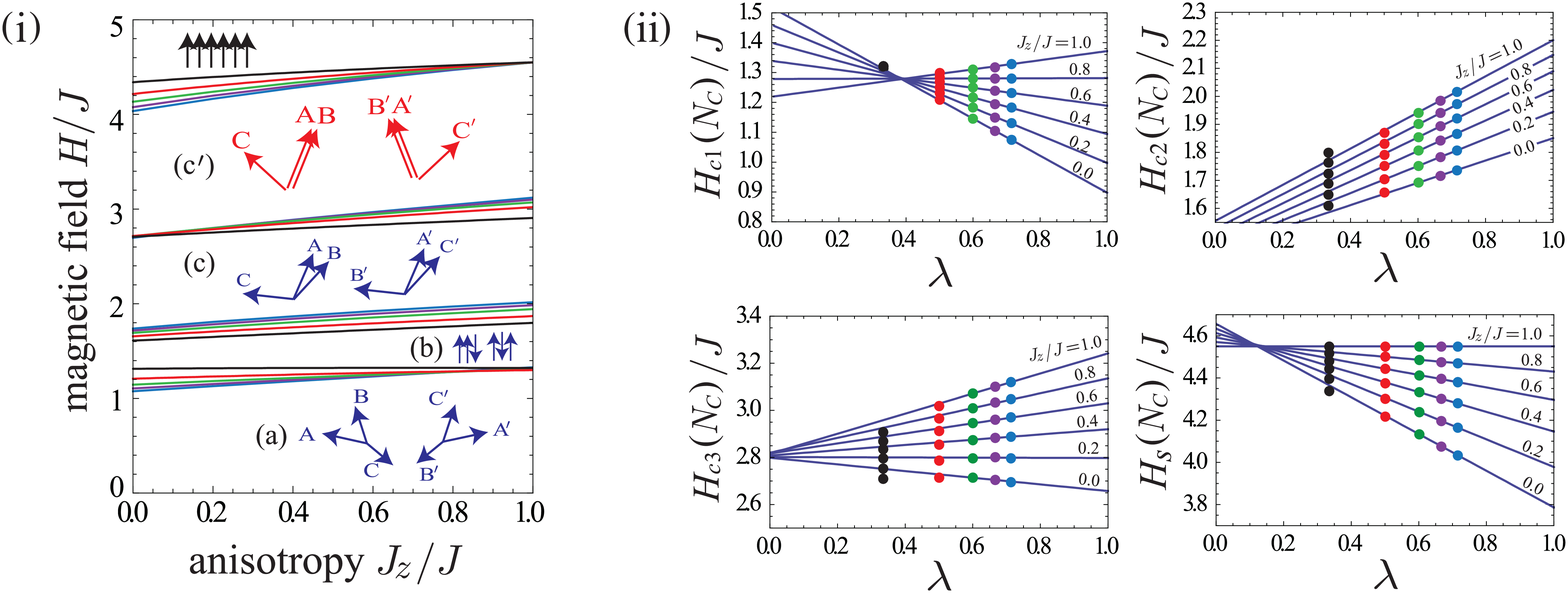}
\caption{\label{figS2}
(i) Phase boundaries obtained with $N_C=3$ (black), 6 (red), 10 (green), 15 (purple), and 21 (blue) for $J^\prime/J =0.025$. (ii) Cluster-size scaling of the data for the phase boundaries. 
}
\end{figure}
\begin{figure}[t]
\includegraphics[scale=0.3]{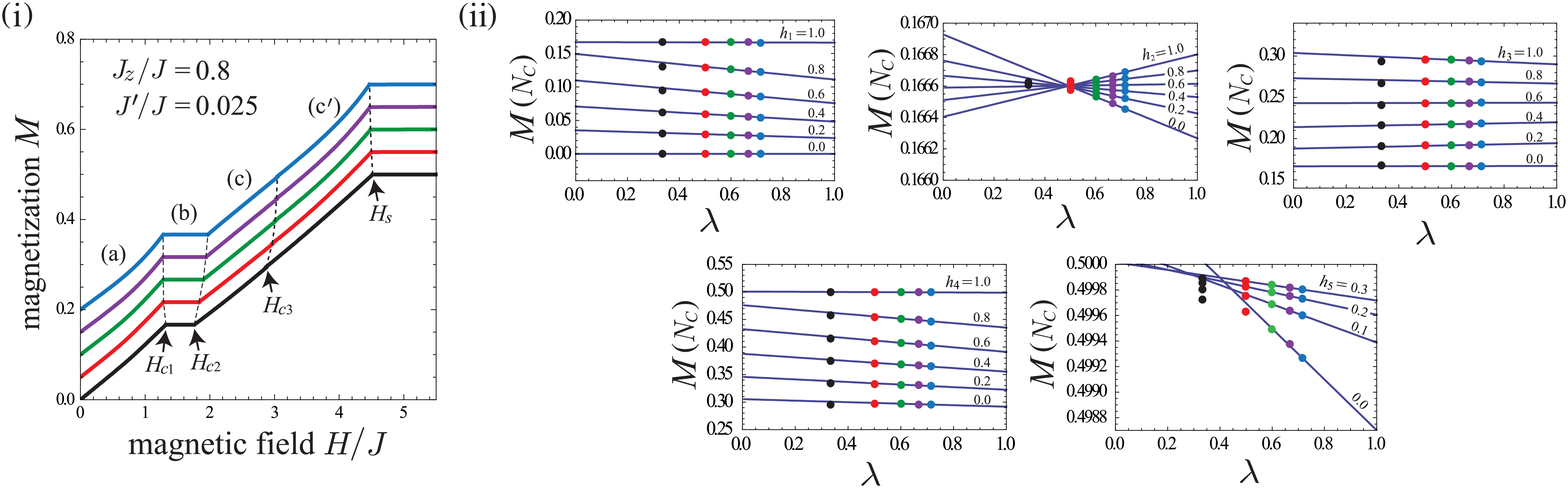}
\caption{\label{figS3}
(i) Ground-state magnetization curves obtained with $N_C=3$ (black), 6 (red), 10 (green), 15 (purple), and 21 (blue) for $J_z/J=0.8$ and $J^\prime/J =0.025$. All the curves apart from the bottom one are vertically shifted by 0.05, 0.1, 0.15, 0.2, respectively, for clarity. (ii) Cluster-size scaling of the data for the ground-state magnetization curve. 
}
\end{figure}
\begin{figure}[t]
\includegraphics[scale=0.55]{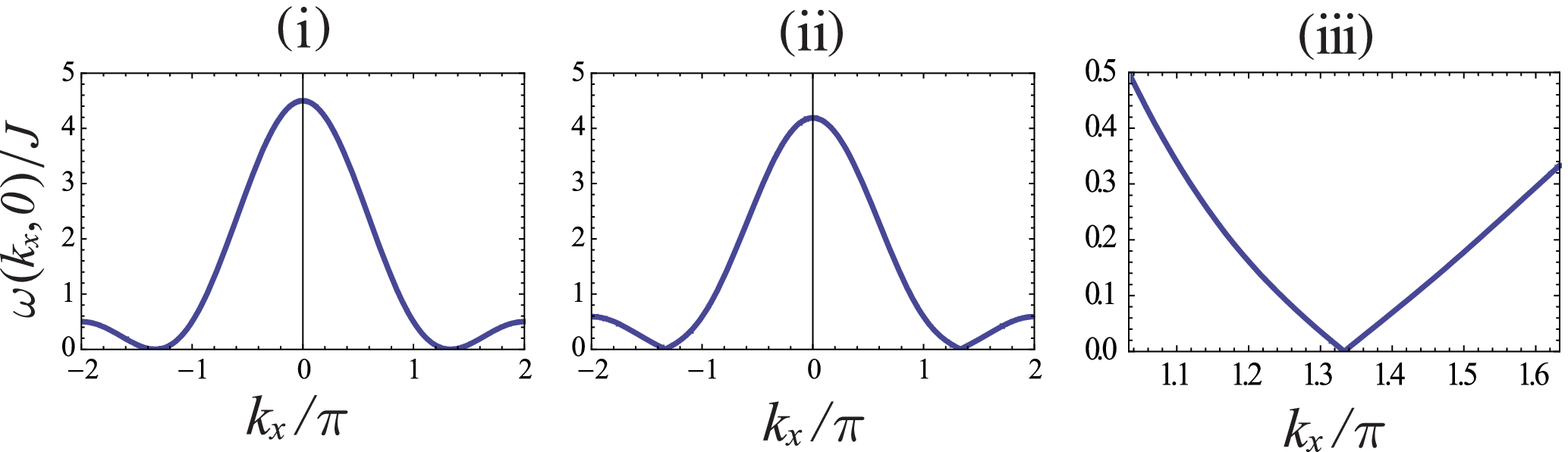}
\caption{\label{figS4}
The spin-wave spectra $\omega(k_x,0)$ at $H=H_s$ for (i) $J_z/J=1$ and (ii) $J_z/J=0.8$. (iii) An enlarged view of (ii).
}
\end{figure}


\begin{thebibliography}{99}
\bibitem{footnote}
We remove a low-field disturbed part of $JdM/dH$ caused by an instrumental problem~\cite{susuki-13}.
\end{thebibliography}
\end{document}